\documentclass[a4paper]{report}
\usepackage[utf8]{inputenc}
\usepackage[T1]{fontenc}
\usepackage{RJournal}
\usepackage{amsmath,amssymb,array}
\usepackage{booktabs}

\usepackage{graphics,epsfig,epsf,psfrag}
\usepackage{color, appendix}
\usepackage{graphicx}
\usepackage{amsmath, amsfonts,amssymb}
\usepackage{epstopdf}
\usepackage{hyperref,url}
\makeatletter \g@addto@macro{\UrlBreaks}{\UrlOrds} \makeatother
\usepackage{enumerate, framed}
\usepackage{graphicx,lscape,rotate,rotating}
\usepackage{algorithmic, algorithm}
\usepackage{fixltx2e}
\usepackage{xcolor}

\def\bSig\mathbf{\Sigma}

\newcommand{\bfbeta}{\mbox{\boldmath $\beta$}}

\newcommand{\bfgamma}{\mbox{\boldmath $\gamma$}}

\newcommand{\z}{\mathbf{z}}

\newcommand{\diag}{\mbox{diag}}

\begin{document}

\sectionhead{Contributed research article}
\volume{XX}
\volnumber{YY}
\year{20ZZ}
\month{AAAA}

\begin{article}
\title{\pkg{coxphMIC}: An R Package for Sparse Estimation of Cox Proportional Hazards Models via 
Approximated Information Criteria}
\author{by Razieh Nabi and Xiaogang Su}

\maketitle

\abstract{
In this paper, we describe an R package named \strong{coxphMIC}, which implements the sparse estimation method for Cox proportional hazards models via approximated information criterion \citep{Su:2016}. The developed methodology is named MIC which stands for ``Minimizing approximated Information Criteria". A reparameterization step is introduced to enforce sparsity while at the same time keeping the objective function smooth. As a result, MIC is computationally fast with a superior performance in sparse estimation. Furthermore, the reparameterization tactic yields an additional advantage in terms of circumventing post-selection inference \citep{Leeb:2005}. The MIC method and its R implementation are introduced and illustrated with the PBC data.
}

\section{Introduction}
Time to event (survival time) is often a primary outcome of interest in many research areas, especially in medical research such as time that takes to respond to a particular therapy, time to death, remission, or relapse. Survival times are typically right skewed and subject to censoring due to study termination, loss of follow ups, or withdrawals. Moreover, covariates may vary by time.

Cox Proportional Hazards (PH) model \citep{Cox:1972} is commonly used to model survival data. Given a typical survival data set that consists of $\{(T_i, \delta_i, \z_i): i = 1, \ldots, n \}$, where $T_i$ is the observed event time, $\delta_i$ is the 0-1 binary censoring indicator, and $\z_i \in \mathbb{R}^p$ is the covariate vector associated with the $i$-th subject, the Cox PH model formulates  the hazard function $h(t|\mathbf{z_i})$ for the $i$th subject as
\begin{equation}
\begin{aligned}
h(t|\mathbf{z_i}) = h_0(t)  \, \exp(\boldsymbol{\beta}^T\ \mathbf{z_i}), \nonumber
\end{aligned}\label{PH-model}
\end{equation}
where $\mathbf{z_i}\in \mathbb{R}^p$ denotes the $p$-dimensional covariate vector associated with subject $i$,  $\boldsymbol{\beta} = (\beta_j)\in \mathbb{R}^p$ is the unknown regression parameter vector, and $h_0(t)$ is the unspecified baseline hazard function. The vector of $\boldsymbol{\beta}$ can be estimated by maximizing the partial log-likelihood \citep{Cox:1975}, which is given by
\begin{equation*}
l(\boldsymbol{\beta}) ~=~ \sum_{i = 1}^n \delta_i \left[\mathbf{z_i}^T\boldsymbol{\beta} - \log
\sum_{i' = 1}^n \left\{I(T_{i'} \geq T_i) \, \exp(\mathbf{z_{i'}}^T \boldsymbol{\beta})
\right\} \right]. \nonumber
\end{equation*}
Let $\widehat{\bfbeta}$ denote the resultant maximum partial likelihood estimator (MPLE).

Since the true $\boldsymbol{\beta}$ is often sparse, we need to look for methods that identify the zero components in $\boldsymbol{\beta}$ and at the same time estimate the nonzero ones. Best subset selection (BSS) and regularization are among two major algorithms used in survival analyses for variable selection. Both are derived from a penalized partial likelihood. Let $\mbox{pen}(\boldsymbol{\beta})$ and $\lambda$ denote the penalty function and penalty parameter, respectively. The general objective function in both of the techniques is as follows:
\begin{equation}
\min_{\boldsymbol{\beta}} ~ - 2 l(\boldsymbol{\beta}) + \lambda \cdot
\mbox{pen}(\boldsymbol{\beta}).  \nonumber \label{penalised-loglik}
\end{equation}

In BSS, the penalty function is set to $\mbox{pen}(\boldsymbol{\beta}) = \sum_{j = 1}^p I\{ \beta_j \neq 0 \}$ (number of nonzero coefficients), and the penalty parameter is fixed as $\lambda = 2$ for AIC \citep{Akaike:1974} or $\lambda = \ln(n_0),$ where $n_0$ is the total number of uncensored failures, with a slight modification of BIC \citep{Vollinsky:2000}. In regularization, the penalty function is set to $\mbox{pen}(\boldsymbol{\beta}) = \sum_{j = 1}^p |\beta_j|$, and the penalty parameter is not fixed and is appropriately chosen. The sparse estimation is reformulated into a continuous convex optimization problem. The optimization of the two techniques is a two-step process. In BSS, one needs to fit every model with the maximum partial likelihood method and then compare the fitted models according to an information criterion such as AIC \citep{Akaike:1974} or BIC \citep{Schwarz:1978}. This makes the BSS infeasible for moderately large $p$. In regularization, one need to solve the objective function for every fixed positive value of $\lambda$ to obtain a regularization path $\{\widetilde{\boldsymbol{\beta}}(\lambda): \, \lambda >0 \},$ and then select the best $\lambda$ according to an information criterion such as AIC  or BIC along the path. Since such a search is only along the regularization path (a one-dimensional curve in $\mathbb{R}^p$), the search space is much reduced and hence, it may not perform as well as the estimator obtained  with BSS, if AIC or BIC is used as the yardstick. Beside the computational burden, both methods face the post-selection inference challenge. A new technique is developed by \citet{Su:2016} on the basis of \citet{Su:2015} for conducting sparse estimation of Cox PH models to help address the aforementioned deficiencies.

\section{The MIC method}
A new method, named MIC for ``Minimizing approximated Information Criteria", is developed to conduct sparse estimation of Cox PH models. MIC borrows strength from both BSS and regularization. The main issue with BSS is the indicator function, $I(\beta \neq 0)$, involved in the $\ell_0$ penalty function, leading to a discrete optimization problem. To overcome this difficulty, MIC proposes to approximate the indicator function by a continuous or smooth unit dent function. One reasonable approximation is the hyperbolic tangent function given by
\begin{equation*}
w(\beta) \,  = \, \tanh (a  \beta^2) \, = \, \frac{\exp \left(a
 \beta^2 \right) - \exp \left(- a
 \beta^2 \right)  }{\exp \left(a  \beta^2 \right) + \exp \left(- a
 \beta^2 \right)}, ~~
\end{equation*}
where $a$ is a nonnegative scale parameter that controls the sharpness of the approximation.

\begin{figure}[h]
\centering
\includegraphics[scale=0.5, angle=0]{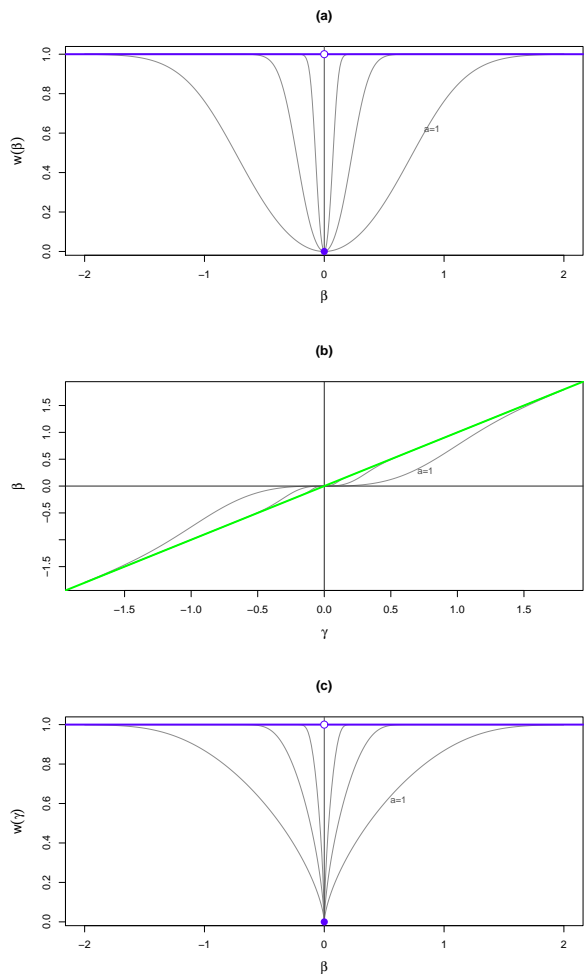}
\caption{MIC penalty and the reparameterization step: (a) the hyperbolic tangent penalty $\tanh(a \beta^2)$ versus $\beta$; (b) $\beta = \gamma \, \tanh(a \gamma^2)$ versus $\gamma$; (c)  $\tanh(a \gamma^2)$ as a penalty function of $\beta$. Three values of $a \in \{1, 10, 100\}$ are illustrated.} \label{fig1}
\end{figure}

As shown in Figure \ref{fig1}(a), $w(\beta)$ provides a smooth approximation to the discrete function $I\{\beta \neq 0\}.$ However, the curve does not have zero as a singular point. If we estimate $\bfbeta$ by minimizing $  - 2 \, l(\bfbeta) + \ln(n_0) \, \sum_{j = 1}^p w(\beta_j),$ we will not obtain sparse estimates. To enforce sparsity, MIC devises a reparameterization step. The reparameterization is based on the decomposition $\beta =  \beta \, I\{ \beta \neq 0\}.$ Set $\gamma = \beta$ and approximate $I\{ \beta \neq 0\}$ with $w(\gamma) = \tanh(a \gamma^2)$. This leads to a reparameterization of $\beta = \gamma w(\gamma)$. The objective function in MIC is given by
\begin{equation}
 Q_n(\bfbeta) ~ = ~ - 2 \, l( \mathbf{W} \gamma) + \lambda_0 \, \mbox{tr}(\mathbf{W}),
 \label{MIC-Qn}
\end{equation}
where the penalty parameter $\lambda_0$ is fixed as $\ln(n_0)$ for BIC \citep{Vollinsky:2000} and matrix $\mathbf{W}$ is $p \times p$ diagonal with diagonal elements $w_j = w(\gamma_j)$ and hence trace $\mbox{tr}(\mathbf{W}) = \sum_{j = 1}^p \tanh(a \gamma_j^2)$. With this notation, it follows that $\bfbeta = \mathbf{W} \bfgamma.$

The above reparameterization offers several important conveniences:
\begin{enumerate}
\item Sparsity now becomes achievable in estimating $\bfbeta$. The penalty $w(\gamma)$ as a function of $\beta = \gamma w(\gamma) = \gamma \tanh(\gamma)$ is a unit dent function that is smooth everywhere except at $\beta = 0$, as shown in Figure \ref{fig1}(c). This is a necessary condition to ensure sparsity as indicated by \citet{Fan:2002}. On this basis, the oracle properties of the MIC estimator $\widetilde{\bfbeta}$ obtained by minimizing $Q_n(\bfbeta)$ in (\ref{MIC-Qn})
$$ \widetilde{\bfbeta} = \arg\min_{\bfbeta} \, Q_n(\bfbeta) = \arg\min_{\bfbeta} - 2 \, l( \bfbeta) + \ln(n_0) \, \sum_{j = 1}^p w(\gamma_j)$$
can be established under regularity conditions. The asymptotic results entails $a_n = O(n)$. For this reason, we fix $a_n = n_0$, the number of non-censored failures. In practice, the empirical performance of MIC is large stable with respect to the choice of $a$, as demonstrated in \citet{Su:2015}. Thus simply fixing $a$ at a reasonably large value (say, $a \geq 10$) could do as well practically.

\item In terms of practical optimization, it is preferable to consider $\bfgamma$ as the decision vector. Namely, we minimize $Q_n(\bfgamma)$ with respect to $\bfgamma$ by treating it as a function of $\bfgamma$. Let $\widetilde{\bfgamma}$ be the resultant MIC estimator of $\bfgamma$
\begin{equation}
 \widetilde{\bfgamma} = \arg\min_{\bfgamma} \, Q_n(\bfgamma) = \arg\min_{\bfgamma} - 2 \, l( \mathbf{W} \gamma) + \ln(n_0) \, \sum_{j = 1}^p w(\gamma_j). \label{MICE-gamma}
\end{equation}
One immediate advantage of doing so is that $Q_n(\bfgamma)$ is smooth in $\bfgamma$ and hence many optimization routines can be applied directly. Since no selection of tuning parameters is involved, MIC is computationally efficient.

\item One consequence of post-selection inference is that no standard error formula is available for zero estimates of $\beta_j$. As depicted in Figure \ref{fig1}(b), $\beta_j$ and $\gamma_j$ have a one-to-one correspondence with $\beta = 0$ iff $\gamma = 0.$ This motivates us to test $H_0:~ \beta_j = 0$ by equivalently testing $H_0:~ \gamma_j = 0.$ The MIC estimator $\widetilde{\bfgamma}$ can be viewed as an M-estimator with smooth objective function $Q_n(\bfgamma)$ and hence standard arguments can be used to make inference.
\end{enumerate}

\section{Implementation in R}

The R package \strong{coxphMIC} implements MIC on the basis of R
package \texttt{survival} \citep{Therneau:2000} and is hosted at
CRAN. Type the following command in
R console in order to install the package:
\begin{example}
> install.packages("coxphMIC")
\end{example}


To summarize, MIC can be simply formulated as the following optimization problem
\begin{equation}
\min_{\bfgamma} ~~ - 2 l(\mathbf{W} \bfgamma) \, + \, \ln(n_0) \sum_{j = 1}^p \tanh(n_0 \gamma_j^2).
\label{MIC}
\end{equation}
Owing to the non-convex nature, a global optimization method is helpful in solving (\ref{MIC}). While other R routines \citep{Mullen:2014} are available, we have found using the \texttt{SANN} method combined with the \texttt{BFGS} method in R function \texttt{optim()} is fast and quite effective.  The simulated annealing (SA) implemented by \texttt{SANN} helps locate a nearly minimum point globally. Then the quasi-Newton BFGS method makes sure that the algorithm stops at a critical point.

There are two functions included in the \strong{coxphMIC} package: an internal function \texttt{LoglikPen()} that computes the partial log-likelihood and a wrapper function \texttt{coxphMIC()} that does the MIC sparse estimation. The function \texttt{coxphMIC()} has the following usage:
\begin{example}
    coxphMIC(formula = Surv(time, status) ~ ., data, method.beta0 = "MPLE",
        beta0 = NULL, theta0 = 1, method = "BIC", lambda0 = NULL, a0 = NULL,
        scale.x = TRUE, maxit.global = 300, maxit.local = 100,
        rounding.digits = 4, zero = sqrt(.Machine$double.eps),
        compute.se.gamma = TRUE, compute.se.beta = TRUE,
        CI.gamma = TRUE, conf.level = 0.95,
        details = FALSE)
\end{example}

We briefly explain some of the important options. The
\texttt{formula} argument is a formula object similar to that in
\texttt{survival}, with the response on the left of the
\texttildelow~ operator being a survival object as returned by the
\texttt{Surv} function, and the terms on the right being
predictors. The arguments \texttt{method.beta0}, \texttt{beta0},
and \texttt{theta0} pertains to the initial starting values. By
default,  the maximum partial likelihood estimator with the option
\texttt{MPLE} is used. Otherwise, one can use the ridge estimator
with option \texttt{ridge}. The \texttt{theta0} corresponds to the
tuning parameter in ridge estimation. User defined starting values
can also be used such as $\bfbeta = \bfgamma = \mathbf{0}$ by
specifying \texttt{beta0}.  By default, the approximated BIC
\citep{Vollinsky:2000} is recommended. However, one can use
\texttt{AIC} \citep{Akaike:1974}. Alternatively, user-specified
penalty is allowed by specifying \texttt{lambda0}. The default
value for $a$ is $n_0$. The option \texttt{maxit.global} allows
for specification of the maximal iteration steps in \texttt{SANN}
while \texttt{maxit.local} specifies the maximal iteration steps
for \texttt{BFGS}. MIC computes the standard errors (SE) for both
$\widetilde{\bfbeta}$ and $\widetilde{\bfgamma}.$ For
$\widetilde{\bfbeta}$, the SE computation is only applicable for
its nonzero components. The option \texttt{maxit.global} asks
whether the user wants to output the confidence intervals for
$\gamma_j$ at the confidence level specified by
\texttt{conf.level} (with 95\% as default).

The output of Function \texttt{coxphMIC()} is an object of S3
class \texttt{coxphMIC}, which is essentially a list of detailed
objects that can be used for other purposes. In particular, the
item \texttt{result} presents the most important results, where
one can see the selected model and inference based on testing
$\bfgamma.$ Two generic functions, \texttt{print} and
\texttt{plot}, are made available for exploring a
\texttt{coxphMIC} object.

\subsection{Other R packages for variable selection in Cox PH models}

Several other R packages are available for variable selection of
Cox PH models. The best subset selection (BSS) is available in the
R Package \strong{glmulti} \citep{Calcagno:2010} with AIC only,
but it is very slow owing to the intensive computation involved.
For large $p$, a stepwise selection procedure could be used as a
surrogate. LASSO \citep{Tibshirani:1997} can be computed via R
Package \strong{glmnet} \citep{Friedman:2010}. \citet{Zhang:2007}
have made their R codes for implementing ALASSO for Cox models
available at
\url{http://www4.stat.ncsu.edu/~hzhang/paper/cox_new.tar}.
But the program was written without resorting well to available R
routines and it takes an unnecessarily long running time. One
alternative way to compute ALASSO is first transform the design
matrix $\mathbf{Z}: = \mathbf{Z} \, \diag(|\widehat{\bfbeta}|)$ so
that LASSO could be applied and then transform the resultant
estimates back. SCAD for Cox PH models \citep{Fan:2002, Fan2010} can be
computed with an earlier version of the R package \strong{SIS}
\citep{Saldana:2016}, but it is no longer available in its current
version. One is referred to Table \ref{tbl3}, which is presented as Table B1 in \citet{Su:2016}, for a comparison study
of these above-mentioned methods. MIC clearly stands out as the
top or among-the-top performer in both sparse estimation and
computing time. 

\begin{table}
\renewcommand{\tabcolsep}{7.pt}
\renewcommand{\arraystretch}{1.2}
\begin{center}
\caption{Comparison of computation time: CPU time (in seconds)
averaged over three runs.}
\vspace{.2in} \centering
\begin{tabular}{ccccccccc}                 \hline
    &       &   Censoring   &   \multicolumn{6}{c}{Method}                 \\ \cline{4-9}
n   &   p   &   Rate    &   Full    &   Stepwise    &   MIC  &
LASSO  &   ALASSO  &   SCAD    \\ \hline
200 &   10  &   25\% &   0.007   &   0.307   &   0.067   &   0.157   &   0.163   &   5.923   \\
    &       &   40\% &   0.000   &   0.320   &   0.060   &   0.150   &   0.170   &   4.900   \\
    &   50  &   25\% &   0.027   &   18.957  &   0.063   &   0.397   &   0.417   &   5.587   \\
    &       &   40\% &   0.027   &   18.107  &   0.057   &   0.453   &   0.480   &   5.480   \\
    &   100 &   25\% &   0.060   &   189.040 &   0.057   &   1.450   &   1.387   &   ---    \\
    &       &   40\% &   0.067   &   181.147 &   0.057   &   2.053   &   1.897   &   ---    \\ \hline
2000    &   10  &   25\% &   0.020   &   0.907   &   0.243   &   0.903   &   0.837   &   415.097 \\
    &       &   40\% &   0.017   &   0.880   &   0.240   &   0.893   &   0.823   &   328.150 \\
    &   50  &   25\% &   0.110   &   81.380  &   0.243   &   1.590   &   1.153   &   ---    \\
    &       &   40\% &   0.093   &   72.887  &   0.237   &   1.613   &   1.163   &   ---    \\
    &   100 &   25\% &   0.333   &   894.607 &   0.223   &   2.383   &   2.103   &   ---    \\
    &       &   40\% &   0.240   &   673.503 &   0.187   &   2.073   &   1.357   &   ---    \\ \hline
\end{tabular}
\label{tbl3}
\end{center}
\end{table}

\section{Examples}
We illustrate the usage of \texttt{coxphMIC} via an analysis of the PBC (primary biliary cirrhosis) data, available from the \strong{survival} package \citep{Therneau:2000}.

\subsection{Data preparation}

To proceed, some minor data preparation is needed. First of all, we want to make sure that the censoring indicator is 0-1 binary.
\begin{example}
> library(survival); data(pbc);
> dat <- pbc; dim(dat);
[1] 418  20
> dat$status <- ifelse(pbc$status == 2, 1, 0)
\end{example}
Next, we explicitly created dummy variable for categorical variables. The \texttt{factor()} function could be used instead. Also, grouped sparsity could be used to handle these dummy variables so that they are either all selected or all excluded. We plan to explore this possibility in future research.
\begin{example}
> dat$sex <- ifelse(pbc$sex == "f", 1, 0)
\end{example}
Another necessary step is to handle missing values. This current version does not automatically treat missings. Here, the listwise deletion is used so that only the 276 subjects with complete records are used for further analysis.
\begin{example}
> dat <- na.omit(dat);
> dim(dat);
[1] 276  20
> head(dat)
  id time status trt      age sex ascites hepato spiders edema bili chol
1  1  400      1   1 58.76523   1       1      1       1   1.0 14.5  261
2  2 4500      0   1 56.44627   1       0      1       1   0.0  1.1  302
3  3 1012      1   1 70.07255   0       0      0       0   0.5  1.4  176
4  4 1925      1   1 54.74059   1       0      1       1   0.5  1.8  244
5  5 1504      0   2 38.10541   1       0      1       1   0.0  3.4  279
7  7 1832      0   2 55.53457   1       0      1       0   0.0  1.0  322
  albumin copper alk.phos    ast trig platelet protime stage
1    2.60    156   1718.0 137.95  172      190    12.2     4
2    4.14     54   7394.8 113.52   88      221    10.6     3
3    3.48    210    516.0  96.10   55      151    12.0     4
4    2.54     64   6121.8  60.63   92      183    10.3     4
5    3.53    143    671.0 113.15   72      136    10.9     3
7    4.09     52    824.0  60.45  213      204     9.7     3
\end{example}
The data set now contains 20 variables. Except \texttt{id}, \texttt{time}, and \texttt{status}, there are a total of 17 predictors.

\subsection{MIC starting with MPLE}

To apply \texttt{coxphMIC}, one simply proceeds in the usual way of using \texttt{coxph} formula. By default, all predictors are standardized; the approximated BIC ($\lambda_0 = \ln(n_0)$ is used with $a = n_0$; and the MPLE is used as the starting point.
\begin{example}
> fit.mic <- coxphMIC(formula = Surv(time, status)~.-id, data = dat, CI.gamma = FALSE)
> names(fit.mic)
 [1] "opt.global" "opt.local"  "min.Q"      "gamma"      "beta"       "VCOV.gamma"
 [7] "se.gamma"   "se.beta"    "BIC"        "result"     "call"
\end{example}
The output of \texttt{coxphMIC} contains the minimized $Q_n$
value, the final estimates of $\bfgamma$ and $\bfbeta$, the
variance-covariance matrix and SE for $\widetilde{\bfgamma}$, SE
for nonzero $\widetilde{\bfbeta}$, BIC value for the final model,
and a summary table \texttt{result}. In order for the user to be
able to inspect the convergence and other detailed info of the
optimization algorithms, we also output two objects
\texttt{opt.global} and \texttt{opt.local}, which result from the
global (\texttt{SANN} by default) and local optimization
(\texttt{BFGS} by default) algorithms.

The output \texttt{fit.mic} is a S3 object of \texttt{coxphMIC}
class. Two generic functions, \texttt{print} and \texttt{plot},
are available. The \texttt{print} function provides a summary
table as below:
\begin{example}
> print(fit.mic)
           beta0   gamma se.gamma  z.stat p.value beta.MIC se.beta.MIC
trt      -0.0622  0.0000   0.1071  0.0000  1.0000   0.0000          NA
age       0.3041  0.3309   0.1219  2.7138  0.0067   0.3309      0.1074
sex      -0.1204  0.0000   0.1086 -0.0002  0.9998   0.0000          NA
ascites   0.0224  0.0000   0.0991  0.0000  1.0000   0.0000          NA
hepato    0.0128  0.0000   0.1259  0.0000  1.0000   0.0000          NA
spiders   0.0460  0.0000   0.1118 -0.0001  1.0000   0.0000          NA
edema     0.2733  0.2224   0.1066  2.0861  0.0370   0.2224      0.0939
bili      0.3681  0.3909   0.1142  3.4237  0.0006   0.3909      0.0890
chol      0.1155  0.0000   0.1181  0.0002  0.9999   0.0000          NA
albumin  -0.2999 -0.2901   0.1248 -2.3239  0.0201  -0.2901      0.1103
copper    0.2198  0.2518   0.1050  2.3986  0.0165   0.2518      0.0868
alk.phos  0.0022  0.0000   0.0837  0.0000  1.0000   0.0000          NA
ast       0.2308  0.2484   0.1128  2.2023  0.0276   0.2484      0.1025
trig     -0.0637  0.0000   0.0858  0.0000  1.0000   0.0000          NA
platelet  0.0840  0.0000   0.1129  0.0000  1.0000   0.0000          NA
protime   0.2344  0.2293   0.1046  2.1917  0.0284   0.2293      0.1022
stage     0.3881  0.3692   0.1476  2.5007  0.0124   0.3692      0.1243
\end{example}
The above results are presented as Table 4 in \citet{Su:2016}. In
this example, MIC started with MPLE given by the first column
named \texttt{beta0}. Columns 2--5 present estimation of
$\bfgamma$ and the hypothesis testing results on $H_0: \gamma_j = 0.$ The estimates of $\bfbeta$ are given in the last two columns.
It can be seen that eight variables are selected in the final
model, which are \texttt{age}, \texttt{edema}, \texttt{bili},
\texttt{albumin}, \texttt{copper}, \texttt{ast}, \texttt{protime},
and \texttt{stage}.

\begin{figure}[h]
\centering
\includegraphics[scale=0.5, angle=270]{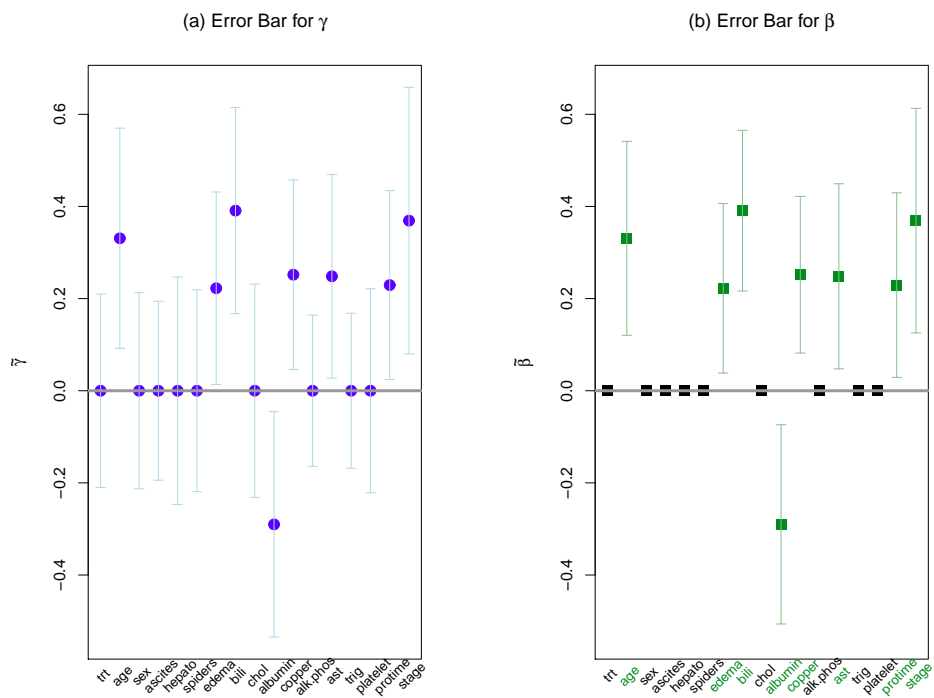}
\caption{Error bar plots for MIC estimates of $\gamma$ in (a) and
$\beta$ in (b). The 95\% confidence intervals (CI) are plotted.
The selected variables are highlighted in green in Panel (b).
\label{fig02}}
\end{figure}

The \texttt{plot} function provides error bar plots based on the
MIC estimator of both $\bfbeta$ and the reparameterized
$\bfgamma:$
\begin{example}
> plot(fit.mic, conf.level = 0.95)
\end{example}
as shown in Figure \ref{fig02}. Essentially, the 95\% confidence
intervals (CI) are plotted. One can modify the confidence level
with the \texttt{conf.level} option. To compare two plots
conveniently, they are made with the same range on the vertical
y-axis. Note that CI is not available for any zero $\beta_j$
estimate in Panel (b), which corresponds to an unselected
variable. Those selected variables are highlighted in green color
in Panel.

\subsection{Multiple starting points}

Trying out multiple starting point is a common strategy in facing global optimization problems. We may consider starting with the $\mathbf{0}$ vector, which corresponds to the null model. Having \texttt{beta0 = }$\mathbf{0}$ is actually the default option if \texttt{method.beta0} is neither `\texttt{MPLE}' nor `\texttt{ridge}' and a specific value for \texttt{beta0} is not given, i.e., setting \texttt{beta0 = NULL}.
\begin{example}
> fit0.mic <- coxphMIC(formula = Surv(time, status)~.-id, data = dat,
+ method = "BIC", scale.x = TRUE, method.beta0 = "zero")

> c(fit.mic$min.Q, fit0.mic$min.Q)
[1] 974.3340 978.1232
\end{example}
We can compare the minimized objective function \texttt{min.Q} to
decide which fitting result is preferable (i.e., the smaller one).
The above result suggests that the fit with MPLE as starting point
remains preferable.

Concerning sparse estimation, the vectors with 0/+1/-1 values
obtained by applying a threshold to the MPLE $|\widehat{\bfbeta}|$
could be reasonable choices for the starting point too, i.e.,
$$ \beta_{0j} ~:=~  \mbox{sgn}(\hat{\beta}_j) \, I\left\{ |\hat{\beta}_j| > c_0\right\},$$
where $c_0 >0$ is a threshold close to 0.
For example, setting $c = 0.06$ yields
\begin{example}
> beta.MPLE <-  fit.mic$result[, 1]
> beta0 <- sign(beta.MPLE)*sign(abs(beta.MPLE) > .06);
> cbind(beta.MPLE, beta0)
      beta.MPLE beta0
 [1,]   -0.0622    -1
 [2,]    0.3041     1
 [3,]   -0.1204    -1
 [4,]    0.0224     0
 [5,]    0.0128     0
 [6,]    0.0460     0
 [7,]    0.2733     1
 [8,]    0.3681     1
 [9,]    0.1155     1
[10,]   -0.2999    -1
[11,]    0.2198     1
[12,]    0.0022     0
[13,]    0.2308     1
[14,]   -0.0637    -1
[15,]    0.0840     1
[16,]    0.2344     1
[17,]    0.3881     1
\end{example}
In the above example, we applied a threshold of 0.06 to the MPLE to obtain a 0/+1/-1 valued vector. To start with this user-supplied starting point, one proceeds as follows.
\begin{example}
> fit1.mic <- coxphMIC(formula = Surv(time, status)~.-id, data = dat,
+ method = "BIC", scale.x = TRUE, method.beta0 = "user-supplied", beta0 = beta0)
> c(fit.mic$min.Q, fit0.mic$min.Q, fit1.mic$min.Q)
[1] 974.3340 978.1232 979.6826
\end{example}
Again, the fitting starting at MPLE seems the best in this example, by giving the
smallest minimized value.

\subsection{Different $a$ values}

We may consider obtaining the regularization path with respect to $a$. According to asymptotic results, $a = O(n)$ is desirable and the recommended value is $a = n_0$ the number of uncensored deaths, which is $n_0 = 111$ in the PBC data under study.

\begin{figure}[h]
\centering
\includegraphics[scale=0.45, angle=270]{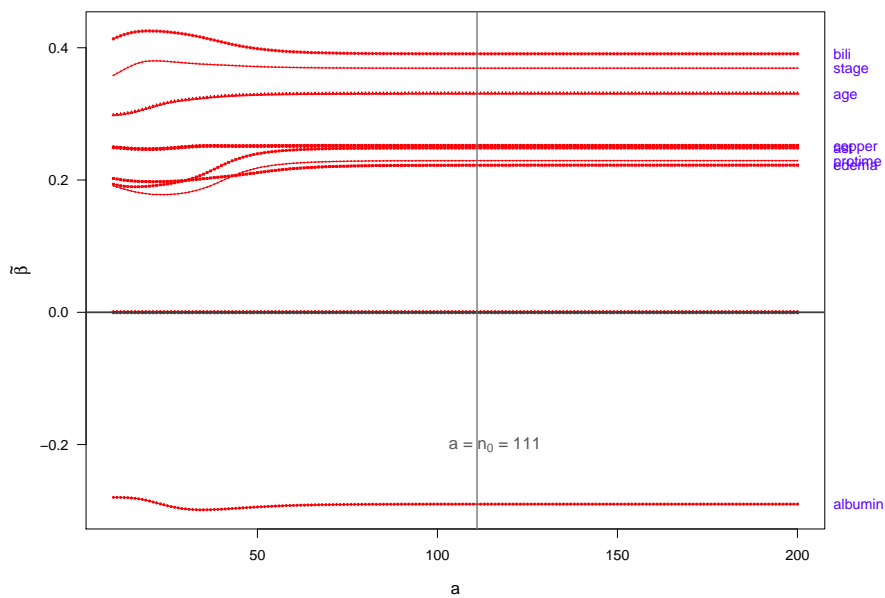}
\caption{The regularization path with respect to $a$ with the PBC
data. The $a$ value varies from 10 to 200. The recommended value
is $a = n_0 = 111.$} \label{fig03}
\end{figure}

We try out a spread of $a$ values that range from 10 to 200, as prescribed by the R object \texttt{A0}.
\begin{example}
> set.seed(818)
> n <- NROW(dat); n0 <- sum(dat$status == 1)
> A0 <- 10:200
> p <- NCOL(dat)-3
> BETA <- matrix(0, nrow = length(A0), ncol = p)   # USE ARRAY
> for (j in 1:length(A0)){
    su.fit <- coxphMIC(formula = Surv(time, status)~.-id, data = dat, a0 = A0[j],
        method = "BIC", scale.x = TRUE)
    BETA[j, ] <- su.fit$beta
  }
> BETA <- as.data.frame(BETA)
> colnames(BETA) <- colnames(dat)[-(1:3)]
> row.names(BETA) <- A0
> head(BETA, n = 5)
   trt    age sex ascites hepato spiders  edema   bili chol albumin copper alk.phos
10   0 0.2983   0       0      0       0 0.2024 0.4135    0 -0.2799 0.2495        0
11   0 0.2987   0       0      0       0 0.2015 0.4159    0 -0.2799 0.2491        0
12   0 0.2992   0       0      0       0 0.2006 0.4181    0 -0.2799 0.2487        0
13   0 0.3000   0       0      0       0 0.1998 0.4200    0 -0.2801 0.2482        0
14   0 0.3009   0       0      0       0 0.1992 0.4216    0 -0.2804 0.2478        0

      ast trig platelet protime  stage
10 0.1937    0        0  0.1912 0.3583
11 0.1924    0        0  0.1895 0.3612
12 0.1914    0        0  0.1878 0.3642
13 0.1906    0        0  0.1862 0.3672
14 0.1901    0        0  0.1847 0.3701
\end{example}
A plot of the regularization path with respect to $a$, as shown in Figure \ref{fig03},
can be obtained as follows:
\begin{example}
> par(mar = rep(5,4), mfrow = c(1,1))
> x.min <- min(A0); x.max <- max(A0)
> plot(x = c(x.min, x.max), y = c(min(BETA), max(BETA)), type = "n",
+    xlab = "a", cex.lab = 1.2, las = 1, ylab = expression(tilde(beta)))
> for (j in 1:ncol(BETA)){
+    lines(x = A0, y = BETA[,j], col = "red", lty = 1, lwd = 1)
+    points(x = A0, y = BETA[,j], col = "red", pch = j, cex = .3)
+    vname <- colnames(BETA)[j]
+    if (abs(BETA[nrow(BETA),j]) > .00001) {
        # text(x.max+5, BETA[nrow(BETA),j], labels = vname, cex = 1, col = "blue")
+        mtext(text = vname, side = 4, line = 0.5, at = BETA[nrow(BETA),j], las = 1,
+            cex = 1, col = "blue", font = 1)
+    }
+ }
> abline(h = 0, col = "gray25", lwd = 2)
> abline(v = n0, col = "gray45", lwd = 1.5)
> text(n0+5, -0.2, expression(paste("a = ", n[0], " = ", 111, sep = "")), cex = 1.2,
+ col = "gray35")
\end{example}
From Figure \ref{fig03}, it can be seen that the regularization
path is essentially flat with respect to $a$, especially for
relatively large $a$ values. This indicates that treating $a$ as a
tuning parameter is unnecessary.

\section{Summary}
The paper presents the \texttt{coxphMIC} package to implement the MIC method for Cox proportional hazards models. Compared to several other competitive methods, MIC has three main advantages by offering a superior empirical performance for it aims to minimize BIC (albeit approximated) without reducing the search space, great computational efficiency since it does not involve selection of any tuning parameter, and a leeway to perform significance testing that is free of the post-selection inference.

\bibliography{nabi-su}
\nocite{*}

\address{Razieh Nabi \\
  Department of Computer Science\\
  Johns Hopkins University\\
  Maryland 21218, USA\\}
  \email{rnabiab1@jhu.edu}

\address{Xiaogang Su\\
  Department of Mathematical Sciences\\
  University of Texas at El Paso\\
  Texas 79968, USA\\}
\email{xsu@utep.edu}

\end{article}

\end{document}